\documentclass[preprint,showpacs,preprintnumbers,amsmath,amssymb]{revtex4}
\usepackage{dcolumn}
\usepackage{bm}
\usepackage[bookmarks,dvipdfm,pdfstartview=FitH]{hyperref}
\usepackage{hypernat}

\begin{document}

\title{Deformed Two-Mode Quadrature Operators in Noncommutative Space}

\author{Hua Wei}
\email{huawei2@eyou.com}
\author{Jiahua Li}%
 \email{lijiahua@mail.hust.edu.cn}
\author{Ranran Fang}
\author{Xiaotao Xie}
\author{Xiaoxue Yang}
\affiliation{Department of Physics, Huazhong University of Science
and Technology, Wuhan 430074, P.R. China}
\date{\today}
\begin{abstract}
Starting from noncommutative quantum mechanics algebra, we
investigate the variances of the deformed two-mode quadrature
operators under the evolution of three types of two-mode squeezed
states in noncommutative space. A novel conclusion can be found and
it may associate the checking of the variances in noncommutative
space with homodyne detecting technology. Moreover, we analyze the
influence of the scaling parameters on the degree of squeezing for
the deformed level and the corresponding consequences.
\end{abstract}

\maketitle

\section{Introduction}
Recently quantum mechanics in noncommutative space are widely
studied. The main motivation for these theories arises from string
theory: the end points of the open strings trapped on an effective
D-brane with a nonzero Neveu-Schwarz two form B-field background
turn out to be noncommuting \cite{connes}. In the presence of the
constant antisymmetric tensor field the momentum operators of the
D-branes have noncommutative structure. The theoretical and
experimental noncommutativity approach in space-space and space-time
field to describe the physics at the Planck scale is widely
discussed
\cite{connes,more,ncqm1,douglas,ncqm2,zjz2004,zjzprl,zjzhth,m-plb,ss,hf,carroll,chaichian,mocioiu}.
Following the limits of M theory and string theory, the
noncommutative gauge theory develops fast, especially noncommutative
space quantum field theory by introducing noncommutative quantum
mechanics (NCQM) \cite{ncqm1,douglas,ncqm2,zjz2004,zjzprl,zjzhth}
has been studied widely, whose applications cover from the
Aharonov-Bohm effect to the quantum Hall effect \cite{m-plb,ss,hf}.
The noncommutative field theories are really challenging because of
their nonlocality, which may have consequences on the ``CPT theorem"
as well as the causality \cite{connes,more,ncqm1,douglas,ncqm2}.

As we all know, near the string scale the space-time appears
noncommutativity, that is, the coordinates became no longer
commutative. The noncommutative structure in space-time can be
introduced in this framework by taking noncommutative coordinates
$\hat{x}_{\mu}$ which satisfy the equation
\begin{equation}
[\hat{x}_{\mu},\hat{x}_{\nu}]=\mathrm{i}\theta_{\mu\nu},
\end{equation}
where $\theta_{\mu\nu}$ is a real and antisymmetric tensor with the
dimensions of length squared.

In recent years, NCQM has focused on some attention
\cite{more,ncqm1,douglas,ncqm2,zjz2004,zjzprl,zjzhth} such as
noncommutative plane, harmonic oscillator, quantum electrodynamics
theory, fractional angular momentum in noncommutative space, and
constraint on quantum gravitational well etc. The dynamics of
charged particle and muliparticle in diverse noncommutative systems
with magnetic field or not are deeply discussed for exploring the
essentially new features of NCQM. In the researching of two
dimensional isotropic harmonic oscillator and testing the spatial
noncommutativity via Rydberg atoms, the authors in Refs.
\cite{zjz2004,zjzprl} from deformed Heisenberg-Weyl Algebra explore
the scheme that the consistent ansatz of commutation relations of
phase variables should simultaneously include space-space
noncommutativity and momentum-momentum noncommutativity. In the
following research \cite{zjzhth}, the author shows that the new type
of boson algebra based on the Bose-Einstein statistics at the
non-perturbation level described by deformed annihilation-creation
operators is determined by the deformed Heisenberg-Weyl algebra
itself, independent of dynamics, where the deformed boson algebra
constitutes a complete and closed algebra. The research illuminates
that the new type of boson algebra including momentum-momentum
noncommutativity is self-contained, and the deformed annihilation
and creation operators of the noncommutative space can be
represented by the undeformed ones via a linear transformation.

Base on the new type of boson algebra, we investigate the variances
of the deformed two-mode quadrature operators under the evolution of
three types of two-mode squeezed states in noncommutative space in
this paper and we get a novel conclusion. We mark the operator in
noncommutative space by using the form adding a hat sign to the
common operator $\hat{O}$ in the present paper, so we can introduce
the corresponding consistent algebra \cite{zjz2004} by
\begin{equation}
\begin{split}\label{1}
[\hat{x}_{i},\hat{x}_{j}]=\mathrm{i}\xi^{2}\theta\epsilon_{ij}\,,\qquad
[\hat{x}_{i},\hat{p}_{j}]=\mathrm{i}\hbar\delta_{ij}\,,\\
[\hat{p}_{i},\hat{p}_{j}]=\mathrm{i}\xi^{2}\eta\epsilon_{ij},\qquad(i,j=a,b),
\end{split}
\end{equation}
where $\theta$ and $\eta$ are the constant parameters, independent
of position and momentum; $\epsilon_{ij}$ is an two-dimensional
antisymmetric unit tensor, $\epsilon_{ab}=-\epsilon_{ba}=1$ and
$\epsilon_{aa}=\epsilon_{bb}=0$. The scaling factor
$\xi=\left(1+\theta\eta/4\hbar^2\right)^{-1/2}$. If
momentum-momentum is commuting, then $\eta=0$, we have $\xi^2=1$,
the NCQM algebra (\ref{1}) reduces to the one which is extensively
discussed in literature for the case that only space-space are
noncommuting.

 Now, let us consider the two-mode light quantum field in noncommutative space.
Following the realization of the Eq.(\ref{1}) by deformed variables
$\hat{x}_{a(b)}$ and $\hat{p}_{a(b)}$ in Ref \cite{zjzhth} and
considering light field environment, we choose the consistent ansatz
of representations of
$\hat{a}(\hat{b})=\frac{\omega}{\sqrt{2}c}\left(\hat{x}_{a(b)}+\frac{\mathrm{i}c^2}{\hbar\omega^2}\hat{p}_{a(b)}\right)$
and
$\hat{a}^\dag(\hat{b}^\dag)=\frac{\omega}{\sqrt{2}c}\left(\hat{x}_{a(b)}-\frac{\mathrm{i}c^2}{\hbar\omega^2}\hat{p}_{a(b)}\right)$,
where $\hat{a}(\hat{b})$ and $\hat{a}^\dag(\hat{b}^\dag)$ are the
annihilation and creation operators in noncommutative space. In
order to maintain Bose-Einstein statistics at the nonperturbation
level \cite{zjz2004,zjzprl,zjzhth}, that is, the operators
$\hat{a}^\dag$ and $\hat{b}^\dag$ should be commuting to keep the
physical meaning, so we have $\eta=\hbar^2\omega^4c^{-4}\theta$ with
the parameter $c$ being velocity of light in vacuum, from
Eqs.(\ref{1}) we can show the commutation relations of
$\hat{a}(\hat{b})$ and $\hat{a}^{\dag}(\hat{b}^{\dag})$
\begin{equation}
\begin{split}
[\hat{a},\hat{a}^{\dag}]=[\hat{b},\hat{b}^{\dag}]=1,\qquad[\hat{a},\hat{b}]=[\hat{a}^{\dag},\hat{b}^{\dag}]=0,\\
[\hat{a},\hat{b}^{\dag}]=[\hat{a}^{\dag},\hat{b}]=\mathrm{i}\xi^2\omega^2c^{-2}\theta.
\end{split}\label{2}
\end{equation}

From the relations $(\ref{1})$ and $(\ref{2})$, we can obtain the
annihilation-creation operators \cite{zjzprl}. For more physical
picture instead of complicated mathematics, we choose
$\tan\phi\equiv\omega^2 c^{-2}\theta/2$ and the scaling factor
$\xi=\left(1+\omega^4 c^{-4}\theta^2/4\right)^{-1/2}=\cos{\phi}$,
this transformation can simplify the calculation by using the
property of the trigonometric function, then we have
\begin{equation}
\begin{split}
\hat{a}&=a\cos\phi+\mathrm{i}b\sin\phi,\qquad\quad\hat{b}=b\cos\phi-\mathrm{i}a\sin\phi,
\\\hat{a}^{\dag}&=a^{\dag}\cos\phi-\mathrm{i}b^{\dag}\sin\phi,\qquad\hat{b}^{\dag}=b^{\dag}\cos\phi+\mathrm{i}a^{\dag}\sin\phi.
\end{split}\label{3}
\end{equation}

The property of single-mode quadrature operators \cite{zjz2004} has
been studied, one finds that variances of single-mode quadrature
operator in one degree of freedom include variances in the other
degree of freedom. The result is of importance and inspire us to go
one step further to clarify the role of the scaling parameters in
the whole process by investigating the features of two-mode
quadrature operators in noncommutative space and we can obtain a
novel conclusion and it may provide a feasible checking scheme to
the noncommutative two-mode squeezed states via homodyne detecting
technology.

The paper is organized as follows. In the following section we show
variances of the two-mode quadrature operators in commutative space.
Then we calculate variances of the two-mode quadrature operators in
noncommutative space and compare the corresponding consequences to
the commutative case, we find variances of the two-mode quadrature
operator in noncommutative space can be written with variances in
commutative space and additional ``correction terms". Finally, we
discuss the corresponding results, analyze the effect of the scaling
parameter on the squeezing degree and clarify the possibility of the
checking scheme to the noncommutative two-mode squeezed states.

\section{Two-mode quadrature operators in commutative space}
In this part, we will show variances of the two-mode quadrature
operators in commutative space. Following the former work
\cite{loudon,fhy}, according to the two-mode quadrature operators in
commutative space
\begin{equation}
\begin{split}
X&=\frac{1}{2\sqrt{2}}\left(a+a^{\dag}+b+b^{\dag}\right),\\
Y&=\frac{1}{\mathrm{i}2\sqrt{2}}\left(a-a^{\dag}+b-b^{\dag}\right),
\end{split}\label{x+y}
\end{equation}
we can study the properties of the two-mode quadrature operators
explicitly using a special effective squeezed state presented by J.
Janszky and A.V. Vinogradov \cite{jj}, which can be achieved through
superposition of coherent states along a straight line on the
$\alpha$ plane. For a single mode of frequency $\omega$, the
electric field operator $E(t)$ is represented as
$E(t)=E_0[a\exp(-i{\omega}t)+a^{\dag}\exp(i{\omega}t)]$, where $a$
and $a^{\dag}$ are the annihilation and creation operators of photon
field. The superposition states is defined by
\begin{equation}
\begin{split}
|\alpha,\pm\rangle&=c_{\alpha\pm}\left(|\alpha\rangle\pm|-\alpha\rangle\right),\\
c_{\alpha\pm}&=\{2[1\pm{\exp(-2{\alpha}^2)}]\}^{-1/2},
\end{split}
\end{equation}
where $|\alpha\rangle$ is the usual coherent state and the algebra
satisfies
$a|\alpha,\pm\rangle={\alpha}c_{\alpha\pm}c_{\alpha\mp}^{-1}|\alpha,\mp\rangle$,
$a^2|\alpha,\pm\rangle={\alpha}^2|\alpha,\pm\rangle$, and
$\langle\alpha,\pm|\alpha,\mp\rangle=0$. For the sake of simplicity,
we have assumed that $\alpha$ is real.

In order to compare the results in commutative space with those in
the noncommutative space, we will first consider the variances of
the two-mode quadrature operators under the evolution of three types
of two-mode squeezed states in commutative space. To achieve this
goal, below let us construct three types of two-mode squeezed
states. From two-mode coherent state $|\alpha\rangle|\beta\rangle$
(the basic algebra
$a|\alpha\rangle|\beta\rangle=\alpha|\alpha\rangle|\beta\rangle$,
$b|\alpha\rangle|\beta\rangle=\beta|\alpha\rangle|\beta\rangle$), we
can define three types of two-mode squeezed states
\cite{zjz2004,jj}, given by
\begin{equation}
\begin{split}
|{\Psi}_1,\pm\rangle\equiv&|\alpha,\pm\rangle|\beta\rangle,\\
|{\Psi}_2,\pm\rangle\equiv&|\alpha,\pm\rangle|\beta,\pm\rangle,\\
|{\Psi}_3,\pm\rangle\equiv&c_{3\pm}\left(|\alpha,\beta\rangle\pm|-\alpha,-\beta\rangle\right),
\end{split}
\end{equation}
where $c^2_{3\pm}=1/\{2[1\pm\exp(-2\alpha^2-2\beta^2)]\}$, and the
corresponding symbol between the equal mark changes synchronously
(e.g., $|{\Psi}_2,+\rangle\equiv|\alpha,+\rangle|\beta,+\rangle$ and
$|{\Psi}_2,-\rangle\equiv|\alpha,-\rangle|\beta,-\rangle$).

For the sake of  describing the squeezing degree of the two-mode
squeezed state, we should take account into the uncertainty relation
via the variance relation
$({\Delta}X)^2={\langle{X^2}\rangle}-{\langle{X}\rangle}^2$ and
$({\Delta}Y)^2={\langle{Y^2}\rangle}-{\langle{Y}\rangle}^2$. For
three different squeezed states, we have obtain the corresponding
results, and the squeezed state appears only in $|\Psi_i,+\rangle$
($i$=1, 2, 3).

First, for the case of $|\Psi_1,+\rangle$, applying the simple
relations
$\langle\Psi_1,+|ab|\Psi_1,+\rangle\!=\!\langle\Psi_1,+|a^{\dag}b^{\dag}|\Psi_1,+\rangle\!=\!0$
and
$\langle\Psi_1,+|a^{\dag}b|\Psi_1,+\rangle\!=\!\langle\Psi_1,+|ab^{\dag}|\Psi_1,+\rangle\!=\!0$,
after some algebra, the variance of the two-mode quadrature operator
can be put into the following form
\begin{equation}
\begin{split}
\left(\Delta{X}\right)^2_{\Psi_{1+}}&=\frac{1}{4}+\frac{\alpha^2}{2[1+\exp(-2\alpha^2)]},\\
\left(\Delta{Y}\right)^2_{\Psi_{1+}}&=\frac{1}{4}-\frac{\alpha^2}{2[1+\exp(2\alpha^2)]},
\end{split}\label{same}
\end{equation}
where the same term $\beta^2/2$ in both
$\langle{X^2}\rangle_{\Psi_{1+}}$, $\langle{Y^2}\rangle_{\Psi_{1+}}$
and ${\langle{X}\rangle}^2_{\Psi_{1+}}$,
${\langle{Y}\rangle}^2_{\Psi_{1+}}$ occurs, so the result shows that
the variance is independent of parameter $\beta$ for
$|\Psi_1,+\rangle$.

Next, for the case of $|\Psi_2,+\rangle$, according to the same
procedure as above, we have the results
\begin{equation}
\begin{split}
\left(\Delta{X}\right)^2_{\Psi_{2+}}&=\frac{1}{4}+\frac{\alpha^2}{2[1+\exp(-2\alpha^2)]}
+\frac{\beta^2}{2[1+\exp(-2\beta^2)]},\\
\left(\Delta{Y}\right)^2_{\Psi_{2+}}&=\frac{1}{4}-\frac{\alpha^2}{2[1+\exp(2\alpha^2)]}
-\frac{\beta^2}{2[1+\exp(2\beta^2)]}.
\end{split}
\end{equation}

Finally, for the case of $|\Psi_3,+\rangle$, according to the
relations
$\langle\Psi_3,+|ab|\Psi_3,+\rangle\!=\!\langle\Psi_3,+|a^{\dag}b^{\dag}|\Psi_3,+\rangle\!=\!\alpha\beta$
and
$\langle\Psi_3,+|a^{\dag}b|\Psi_3,+\rangle\!=\!\langle\Psi_3,+|ab^{\dag}|\Psi_3,+\rangle\!=\!\alpha\beta\frac{c^{2}_{3+}}{c^{2}_{3-}}$,
 the variances of $X$ and $Y$ are given by
\begin{equation}
\begin{split}
\left(\Delta{X}\right)^2_{\Psi_{3+}}&=\frac{1}{4}
+\frac{(\alpha+\beta)^2}{2[1+\exp(-2\alpha^2-2\beta^{2})]},\\
\left(\Delta{Y}\right)^2_{\Psi_{3+}}&=\frac{1}{4}
-\frac{(\alpha+\beta)^2}{2[1+\exp(2\alpha^2+2\beta^{2})]}.
\end{split}
\end{equation}

\section{Deformed two-mode quadrature operators in noncommutative space}
In this section let us show what is the situation for the two-mode
quadrature operators in the noncommutative space. Using the
relations (\ref{3}) and (\ref{x+y}), we can construct the deformed
two-mode quadrature operators as follows
\begin{equation}
\begin{split}
\hat{X}\!\equiv\frac{1}{2\sqrt{2}}\left(\hat{a}+\hat{a}^{\dag}+\hat{b}+\hat{b}^{\dag}\right)
&=X\cos\phi+Y^{ab}\sin\phi,\\
\hat{Y}\!\equiv\frac{1}{\mathrm{i}2\sqrt{2}}\left(\hat{a}-\hat{a}^{\dag}+\hat{b}-\hat{b}^{\dag}\right)
&=Y\cos\phi+X^{ab}\sin\phi,\\
[\hat{X},\hat{Y}]=\frac{\mathrm{i}}{2}\,,
\end{split}\label{xy}
\end{equation}
where we have introduced the definitions $Y^{ab}=(Y^a-Y^b)/\sqrt{2}$
and $X^{ab}=(X^b-X^a)/\sqrt{2}$ for the sake of convenience. The
quantities $X^a$, $X^b$, $Y^a$ and $Y^b$ are the corresponding
a(b)-mode quadrature operators, denoted by
$X^a=\frac{1}{2}(a+a^{\dag}),Y^a=\frac{1}{2\mathrm{i}}(a-a^{\dag})$
and
$X^b=\frac{1}{2}(b+b^{\dag}),Y^b=\frac{1}{2\mathrm{i}}(b-b^{\dag})$,
which are frequently used in the squeezed quantum field. The
expressions of $X$ and $Y$ are identical to Eqs.\,(\ref{x+y}). It is
pointed out that the deformed quadrature operators satisfy the same
commuting relation as the case in commutative space.

In what follows we concentrate on the variances of deformed two-mode
quadrature operators. Inserting Eqs.(\ref{xy}) into the relations
$(\Delta{\hat{X}})^2=\langle{\hat{X}^2}\rangle-\langle{\hat{X}}\rangle^2$
and
$(\Delta{\hat{Y}})^2=\langle{\hat{Y}^2}\rangle-\langle{\hat{Y}}\rangle^2$,
we have
\begin{eqnarray}
\begin{split}
(\Delta{\hat{X}})^2&=\left\langle\left(X\cos\phi+Y^{ab}\sin\phi\right)^2\right\rangle
-\left\langle X\cos\phi+Y^{ab}\sin\phi\right\rangle^2,\\
(\Delta{\hat{Y}})^2&=\left\langle\left(Y\cos\phi+X^{ab}\sin\phi\right)^2\right\rangle
-\left\langle Y\cos\phi+X^{ab}\sin\phi\right\rangle^2,
\end{split}\label{li}
\end{eqnarray}
where
$\cos^2\phi=\xi^2=\left(1+{\omega^4c^{-4}\theta^2}/4\right)^{-1}$
and $\sin^2\phi=1-\xi^2$ are the $\theta$-dependent experiment
parameters \cite{ncqm1,zjzprl,carroll}.

If we expand equations above and do some rearrangement, we can
arrive at a novel consequence
\begin{eqnarray}
\begin{split}
(\Delta{\hat{X}})^2&=(\Delta X)^2+\left[(\Delta Y^{ab})^2-(\Delta X)^2\right]\sin^2\phi+C_{\Delta{\hat{X}}},\\
(\Delta{\hat{Y}})^2&=(\Delta Y)^2+\left[(\Delta X^{ab})^2-(\Delta
Y)^2\right]\sin^2\phi+C_{\Delta{\hat{Y}}},
\end{split}\label{expand}
\end{eqnarray}
where $C_{\Delta{\hat{X}}}$ and $C_{\Delta{\hat{Y}}}$ represent the
cross terms of the Eqs.(\ref{li}), the specific expression are
$C_{\Delta{\hat{X}}}=(\langle XY^{ab}+Y^{ab}X\rangle-2\langle
X\rangle\langle Y^{ab}\rangle)\sin\phi\cos\phi$ and
$C_{\Delta{\hat{Y}}}=(\langle YX^{ab}+X^{ab}Y\rangle-2\langle
Y\rangle\langle X^{ab}\rangle)\sin\phi\cos\phi$, respectively. The
calculation process become compact instead of complicated via
introducing the trigonometric function transformation.

From Eqs.(\ref{expand}), we can find variances of the two-mode
quadrature operator in noncommutative space can be written with
variances in commutative space and additional ``correction terms",
the scaling parameter has no contribute to the first term in
Eqs.(\ref{expand}). It it inspiring, because the quadrature variance
of squeezed state can be measured directly via homodyne detection
technology \cite{homodyne}. This is a very interesting result for
the influence of the noncommutative space to the two-mode squeezing
field and it can provide a novel checking scheme to the
noncommutative two-mode squeezed states in a simple way. Following
the universal equations above, we can get the corresponding results
for different type of the two-mode squeezing light field in
noncommutative space, respectively.

To begin with, for the case of $|\Psi_1,+\rangle$, after some
conversion we have
$(C_{\Delta\hat{X}})_{\Psi_1+}=(C_{\Delta\hat{Y}})_{\Psi_1+}=0$,
$(\Delta Y^{ab})^2_{\Psi_1+}=1/4-\alpha^2/\{2[1+\exp(2\alpha^2)]\}$
and $(\Delta
X^{ab})^2_{\Psi_1+}=1/4+\alpha^2/\{2[1+\exp(-2\alpha^2)]\}$.
Substituting the above results into Eq.\,(\ref{li}), the
 corresponding variances follow
\begin{equation}
\begin{split}
(\Delta\hat{X})^2_{\Psi_1+}=(\Delta X)^2_{\Psi_1+}-\frac{\alpha^2}{2}(1-\xi^2),\\
(\Delta\hat{Y})^2_{\Psi_1+}=(\Delta
Y)^2_{\Psi_1+}+\frac{\alpha^2}{2}(1-\xi^2),
\end{split}\label{P1}
\end{equation}
we can find the only modification from noncommutative space is a
factor $\alpha^2(1-\xi^2)/2$ here comparing to Eq.(\ref{same}) for
the state vector, here the cross term in Eq.(\ref{expand}) is zero
and it is same to state vectors else.

Next, for the case of $|\Psi_2,+\rangle$, after some algebra we have
$(C_{\Delta\hat{X}})_{\Psi_2+}=(C_{\Delta\hat{Y}})_{\Psi_2+}=0$,
$(\Delta
Y^{ab})^2_{\Psi_2+}=1/4-\alpha^2/\{2[1+\exp(2\alpha^2)]\}-\beta^2/\{2[1+\exp(2\beta^2)]\}$
and $(\Delta
Y^{ab})^2_{\Psi_2+}=1/4+\alpha^2/\{2[1+\exp(-2\alpha^2)]\}+\beta^2/\{2[1+\exp(-2\beta^2)]\}$.
 As such, we can obtain the corresponding results
\begin{equation}
\begin{split}
(\Delta\hat{X})^2_{\Psi_2+}=(\Delta X)^2_{\Psi_2+}-\frac{(\alpha^2+\beta^2)}{2}(1-\xi^2)\\
(\Delta\hat{Y})^2_{\Psi_2+}=(\Delta
Y)^2_{\Psi_2+}+\frac{(\alpha^2+\beta^2)}{2}(1-\xi^2)
\end{split}\label{P2}
\end{equation}
the only modification from noncommutative space is a factor
$(\alpha^2+\beta^2)(1-\xi^2)/2$.

And for the case of $|\Psi_3,+\rangle$, making full use of
$(C_{\Delta\hat{X}})_{\Psi_3+}=(C_{\Delta\hat{Y}})_{\Psi_3+}=0$,
$(\Delta
Y^{ab})^2_{\Psi_3+}=1/4-(\alpha-\beta)^2/\{2[1+\exp(\alpha^2+\beta^2)]\}$
and $(\Delta
X^{ab})^2_{\Psi_3+}=1/4+(\alpha-\beta)^2/\{2[1+\exp(-\alpha^2-\beta^2)]\}$
we can obtain corresponding variances
\begin{equation}
\begin{split}
(\Delta\hat{X})^2_{\Psi_3+}=(\Delta X)^2_{\Psi_3+}-\left\{\frac{\alpha^2+\beta^2}{2}+\frac{\alpha\beta[1-\exp(-2\alpha^2-2\beta^2)]}{1+\exp(-2\alpha^2-2\beta^2)}\right\}(1-\xi^2),\\
(\Delta\hat{Y})^2_{\Psi_3+}=(\Delta
Y)^2_{\Psi_3+}+\left\{\frac{\alpha^2+\beta^2}{2}+\frac{\alpha\beta[1-\exp(-2\alpha^2-2\beta^2)]}{1+\exp(-2\alpha^2-2\beta^2)}\right\}(1-\xi^2).
\end{split}\label{P3}
\end{equation}
the only modification from noncommutative space is a term depending
on the scaling parameter too.

Equations (\ref{P1})-(\ref{P3}) are the main results for the
two-mode quadrature operators under the evolution of three types of
two-mode squeezed states in noncommutative space, they all come from
the universal Equations (\ref{expand}). We can compare these results
to the commutative case and analyze the influence of the scaling
parameters on the degree of squeezing for the deformed level and the
corresponding consequences.

\section{Discussions and conclusions}
There are different bounds on the parameter $\theta$ set by
experiments. Although $\theta$ is surely small, the existing
experiments \cite{ncqm1,zjzprl,carroll,chaichian,mocioiu}
demonstrate the approximate order of magnitude. The low-energy
test\cite{carroll} of Lorentz invariance place bounds on
noncommutative energy scale of order 10 Tev. Measurements of the
Lamb shift \cite{chaichian} give a weaker bound. Far stronger bounds
arise if one considers potential noncommutative effects in strong
interactions\cite{mocioiu}. For the scaling factor
$\xi^2=\left(1+\omega^4c^{-4}\theta^2/4\right)^{-1}$, from the
specific experimental parameter we can analyze the scaling factor in
detail. On the other hand, detection and measurement of squeezed
states via homodyne detection is a proven technology
\cite{homodyne}. It may be a feasible scheme to study NCQM utilizing
the auxiliary deformed two-mode squeezed state. In order to
associate the variances with the experiment parameter $\theta$, we
expand the scaling factor $\xi^2$ with respect to $\theta$ using
Taylor power series
\begin{equation}
\begin{split}
\xi^2=1-A\theta^2+O(\theta^{4}),
\end{split}\label{p7}
\end{equation}
where $A=\omega^4c^{-4}/4$, and we only retain the terms up to the
second order of the parameter $\theta$. Inserting the Taylor
expansion (\ref{p7}) into Eqs.\,(\ref{P1})-(\ref{P3}), the influence
of the noncommutativity on the two-mode quadrature operators can be
determined by
\begin{equation}
\begin{split}\label{P4}
(\Delta\hat{X})^2_{\Psi_1+}=(\Delta{X})^2_{\Psi_{1+}}-\frac{\alpha^2}{2}A\theta^2,\\
(\Delta\hat{Y})^2_{\Psi_1+}=(\Delta{Y})^2_{\Psi_{1+}}+\frac{\alpha^2}{2}A\theta^2,
\end{split}
\end{equation}
\begin{equation}
\begin{split}\label{P5}
(\Delta\hat{X})^2_{\Psi_2+}=(\Delta{X})^2_{\Psi_{2+}}-\frac{\alpha^2+\beta^2}{2}A\theta^2,\\
(\Delta\hat{Y})^2_{\Psi_2+}=(\Delta{Y})^2_{\Psi_{2+}}+\frac{\alpha^2+\beta^2}{2}A\theta^2,
\end{split}
\end{equation}
\begin{equation}
\begin{split}\label{P6}
(\Delta\hat{X})^2_{\Psi_3+}=(\Delta{X})^2_{\Psi_{3+}}-\left\{\frac{\alpha^2+\beta^2}{2}+\frac{\alpha\beta[1-\exp(-2\alpha^2-2\beta^2)]}{1+\exp(-2\alpha^2-2\beta^2)}\right\}A\theta^2,\\
(\Delta\hat{Y})^2_{\Psi_3+}=(\Delta{Y})^2_{\Psi_{3+}}+\left\{\frac{\alpha^2+\beta^2}{2}+\frac{\alpha\beta[1-\exp(-2\alpha^2-2\beta^2)]}{1+\exp(-2\alpha^2-2\beta^2)}\right\}A\theta^2.
\end{split}
\end{equation}

From Eqs. (\ref{P4})-(\ref{P6}), we can arrive at the conclusion
that the degree of the squeezing in noncommutative space becomes
attenuated explicitly in Y direction and the degree of the squeezing
in X direction becomes enhanced, as compared to that in commutative
space. When $\theta=0$, it is straightforward to show from Eqs.
(\ref{P1})-(\ref{P3}) or Eqs. (\ref{P4})-(\ref{P6}) that the
variances in noncommutative space turn back into the ones in
commutative space, which is in good agreement with the commutation
relations (\ref{2}). Alternatively, when the noncommutative system
is determinate, the variances of the two-mode quadrature operators
under the evolution of three types of two-mode squeezed states can
be written in the form of the simple addition or subtraction with
the variances in commutative condition and some ``correction terms"
connected with experiment parameter $\theta$.

In summary, starting from noncommutative quantum mechanics algebra,
we investigate the variances of the two-mode quadrature operators
under the evolution of three types of two-mode squeezed states in
commutative and noncommutative space, and point out the difference
between them. In addition, we analyze the influence of the scaling
parameters on the degree of the squeezing for the deformed level and
illuminate the corresponding conclusions, which may be helpful for
testing the uncertainty relation and the essential features of the
noncommutative space in corresponding experiments. Our results are
of theoretical interest in understanding the characteristics of the
two-mode squeezing light fields in noncommutative space, and  have
potential application in searching new schemes for the relating
experimental testing of noncommutative space.

\begin{acknowledgments}
The authors would like to thank Prof. Ying Wu for valuable
discussions and useful comments. The work is  supported in part by
the National Natural Science Foundation of China Under Grant Nos.
90503010 and 10575040, and by National Basic Research Program of
China 2005CB724508.
\end{acknowledgments}

\end{document}